\documentclass[%
 reprint,
 amsmath,amssymb,
 aps,
]{revtex4-2}

\usepackage{graphicx}
\usepackage{dcolumn}
\usepackage{bm}

\usepackage{ulem}

\newcommand{\updated}[1]{\textcolor{black}{#1}}

\usepackage[utf8]{inputenc}
\usepackage{amsmath}
\usepackage{layout}
\usepackage{mathtools}

\usepackage{xcolor}
\usepackage{graphicx}
\usepackage{amssymb}
\usepackage{siunitx}
\usepackage{float}
\usepackage{amsthm}
\usepackage{tikz-cd}
\usepackage{hyperref}
\usepackage{bbm}
\hypersetup{
    colorlinks = true,
    linkcolor = violet,
    linkbordercolor = {white}
}

\newcommand{\R}{\mathbb{R}}

\newcommand{\Z}{\mathbb{Z}}
\newcommand{\C}{\mathbb{C}}

\newcommand{\seq}{\subseteq}

\newcommand{\ket}[1]{\vert #1 \rangle}
\newcommand{\bra}[1]{\langle #1 \vert}
\newcommand{\ketbra}[2]{\ket{#1} \bra{#2}}
\newcommand{\bkop}[3]{\langle #1 \vert #2 \vert #3 \rangle}
\newcommand\norm[1]{\left\lVert#1\right\rVert}

\newcommand{\twoMat}[4]{\begin{pmatrix} #1 & #2 \\ #3 & #4 \end{pmatrix}}

\newcommand{\mb}[1]{\mathbf{#1}}

\newcommand{\br}[1]{\left( #1 \right)}

\renewcommand{\exp}[1]{\text{exp}\br{#1}} 
\newcommand{\set}[1]{\{ #1 \}} 

\DeclareMathOperator{\SWAP}{SWAP}
\DeclareMathOperator{\CNOT}{CNOT}

\newcommand{\bounds}[2]{\bigg\rvert_{#1}^{#2}}
\newcommand{\boudns}[2]{\bounds} 

\renewcommand{\a}{\alpha}
\renewcommand{\b}{\beta}

\renewcommand{\d}{\delta}
\newcommand{\D}{\Delta}
\newcommand{\e}{\epsilon}

\renewcommand{\l}{\lambda}
\newcommand{\m}{\mu}

\newcommand{\p}{\phi}
\renewcommand{\r}{\rho}
\newcommand{\s}{\sigma}
\renewcommand{\t}{\tau}
\renewcommand{\th}{\theta}

\newcommand{\W}{\Omega}

\newcommand{\bth}{{\boldsymbol\th}}

\newcommand{\CH}{\mathcal{H}}

\newcommand{\CL}{\mathcal{L}}

\DeclareMathOperator{\KL}{KL}
\DeclareMathOperator{\QKL}{QKL}

\DeclareMathOperator{\Unif}{Unif}

\begin{document}

\preprint{APS/123-QED}


\title{Learning quantum symmetries with interactive quantum-classical variational algorithms
 }

\author{Jonathan Z. Lu}
\email{jlu@college.harvard.edu}
\author{Rodrigo Araiza Bravo}%
\thanks{This symbol denotes equal author contribution.}
\author{Kaiying Hou} 
\thanks{This symbol denotes equal author contribution.}
\author{\\Gebremedhin A. Dagnew}
\altaffiliation[Presently at ]{1QB Information Technologies, Inc.}

\author{Susanne F. Yelin} 
\email{syelin@g.harvard.edu}
\author{Khadijeh Najafi}
\email{knajafi@ibm.com}
\altaffiliation{IBM Quantum, IBM T.J. Watson Research Center, Yorktown Heights, NY 10598 USA}
\affiliation{%
 Department of Physics, Harvard University, Cambridge, MA
}%

\date{\today}

\begin{abstract}
A symmetry of a state $\ket{\psi}$ is a unitary operator of which $\ket{\psi}$ is an eigenvector. When $\ket{\psi}$ is an unknown state supplied by a black-box oracle, the state's symmetries provide key physical insight into the quantum system; symmetries also boost many crucial quantum learning techniques. In this paper, we develop a variational hybrid quantum-classical learning scheme to systematically probe for symmetries of $\ket{\psi}$ with no a priori assumptions about the state. 
This procedure can be used to learn various symmetries at the same time. In order to avoid re-learning already known symmetries, we introduce an interactive protocol with a classical deep neural net. The classical net thereby regularizes against repetitive findings and allows our algorithm to terminate empirically with all possible symmetries found. Our scheme can be implemented efficiently on average with non-local SWAP gates; we also give a less efficient algorithm with only local operations, which may be more appropriate for current noisy quantum devices. We simulate our algorithm on representative families of states, \updated{including cluster states and ground states of Rydberg and Ising Hamiltonians}. We also find that the numerical query complexity scales well with qubit size. 
\end{abstract}
\maketitle

\section{\label{sec:intro}Introduction}
\begin{figure*}[ht!]
    \centering
    \includegraphics[scale=0.55]{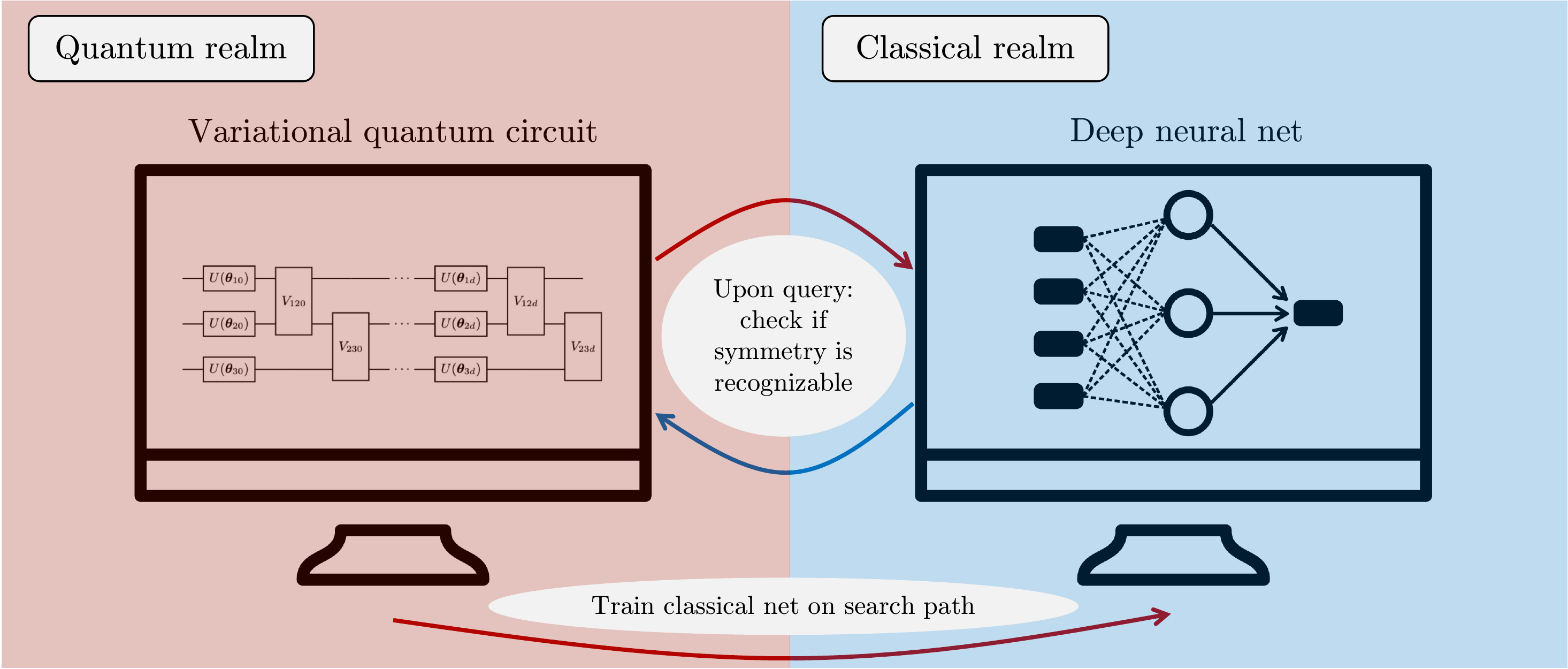}
    \caption{Schematic overview of symmetry learning scheme. A variational quantum circuit minimizes a loss function to generate symmetries, at the same time training a classical neural network to recognize the path of potential symmetry operators it searches. The classical net then alerts the quantum circuit when it is searching for symmetries similar to those already found, so that the quantum circuit can redirect its search accordingly.}
    \label{fig:Schematic}
\end{figure*}

Symmetries are of ubiquitous importance across physics, relating intimately to conservation laws and guiding the formulation of physical theories. The search for symmetries in a given physical system is a central problem in the field, and until recently, has remained a purely analytical and model-driven task. However, developments in machine learning have now opened the door for data-driven approaches. Classically, recent studies have used learning algorithms to discover conserved quantities by using the system's equation of motion~\cite{Mototake_2021,Wetzel2020,Seungwoong_2021,liu2022ai,Liu_AI2} or by clustering methods~\cite{krippendorf2020detecting}. In the quantum setting, classical neural networks have been successfully trained to classify states of matter~\cite{broecker2017machine, dong2019machine, carleo2017solving}, including symmetry-protected topological ones~\cite{zhang2020interpreting, deng2017machine, kaming2021unsupervised}. Quantum algorithms binarily testing for symmetries of certain Hamiltonians and for specific quantum symmetries (such as bosonic exchange) have too been constructed~\cite{laborde2022ham,laborde2021testing}. Knowledge of a given state's symmetries significantly boosts the power of many fundamental quantum learning schemes~\cite{meyer2022exploiting,larocca2022group,zheng2021speeding}. Along these lines, it has been shown that knowledge of symmetries can bootstrap the quantum approximate optimization algorithm (QAOA)~\cite{zhou2022qaoa}. Symmetry-abiding neural networks can also be used as a variational ansatz for states of lattice models~\cite{luo2021gaugeA, luo2021gaugeB}.

Beyond their intrinsic importance, quantum symmetries also encode key information about the underlying quantum state.
Specifically, when a state $\ket{\psi}$ is unknown, efficient extraction of information about $\ket{\psi}$ from limited data is of great interest across both physics and theoretical computer science. Full information extraction, however, is exponentially costly with the system size, requiring complete state tomography~\cite{paris2004quantum}. While novel approaches circumventing this issue have appeared---including basis-enhanced Born machines, classical machine learning, and shadow tomography---comparatively little attention has been paid to the direct and systematic learning of quantum symmetries, which also serve to characterize a state~\cite{carleo_solving_2017, torlai_integrating_2019, torlai_neural-network_2018, carrasquilla_reconstructing_2019, wang_scalable_2020, carrasquilla_neural_2021, cha_attention-based_2022,huang_provably_2021,BEBM_Gomez2022}. In this article, we take a first step in that direction. 

Define a quantum symmetry of $\ket{\psi}$ to be a unitary $U$ such that 
\begin{align}
    U \ket{\psi} = e^{i \th} \ket{\psi} ;
\end{align}
that is, an operator $U$ for which $\ket{\psi}$ is an eigenvector. 
For maximal generality, we shall assume that a black-box oracle machine prepares copies of an unknown state $\ket{\psi}$; our goal is to efficiently express the set of symmetries of $\ket{\psi}$. To find the symmetries of $\ket{\psi}$ \footnote{
We restrict ourselves to pure states $\ket{\psi}$ and system-isolated operators (i.e. acting only on $\ket{\psi}$) for simplicity. However, The algorithm generalizes readily to a density matrix and Krauss operator-sum formalism.}, we propose a hybrid quantum-classical algorithm consisting of a variational quantum circuit interacting with a classical neural network. Under the oracle model, we assume no prior information about $\ket{\psi}$, so that our algorithm is agnostic to the input state whose symmetries it learns. The quantum circuit generates symmetries of $\ket{\psi}$ while simultaneously training the classical net to ``remember" the history of symmetries already found. In turn, the classical net alerts the quantum circuit when it parameterizes an operator close to a previously-found symmetry, allowing the quantum circuit to redirect its search and thereby avoid repeatedly generating similar symmetries. By both generating symmetries and keeping track of known symmetries, our algorithm continuously finds new symmetries. Thus, we observe empirically that the learning procedure eventually terminates. The overall structure of quantum-classical interaction is shown in Fig.~\ref{fig:Schematic}.

In Section~\ref{sec:theory}, we discuss the full details of the symmetry learning scheme and place it within the context of noisy near-term quantum hardware. In Section~\ref{sec:results}, we demonstrate the algorithm on three representative families of quantum states.
We also benchmark the scalability of the algorithm and its robustness to noise. Finally, in Section~\ref{sec:conclusion} we summarize our findings and discuss some applications of symmetry learning.

\section{\label{sec:theory}Hybrid Learning Scheme}
By analogy to an experimental setting, we assume access to copies of an unknown state $\ket{\psi}$ and the task is to discover the set of its symmetries.  Since the closeness of an operator to a symmetry can be quantified in a manner similar to metrics of state overlap, symmetry learning has a natural interpretation as an optimization problem, given as
\begin{align}
    S[\ket{\psi}] = \set{V \in \mathrm{U}(2^L) \;:\; |\bkop{\psi}{V}{\psi}|^2 = 1} ,
\end{align} 
where $\mathrm{U}(2^L)$ is the unitary matrix Lie group of dimension $2^L$ (for $L$ qubits). Symmetries can be filtered into a collection of sets by fixing a universal parameterized quantum circuit (PQC) family $C_{L, D}(\bth)$ of depth $D$ on $L$ qubits and parameters $\bth$. Define $S_D[\ket{\psi}]$ as the collection of symmetries of $\ket{\psi}$ representable by $C_{L, D}(\bth)$. By universality, every symmetry is contained in $S_{O(\exp{L})}[\ket{\psi}]$ and 
\begin{align}
    S_0[\ket{\psi}] \seq S_1[\ket{\psi}] \seq \cdots S_{O(\exp{L})}[\ket{\psi}] .
\end{align} 
We formally define the symmetry learning problem as the discovery and classification of all symmetries of $\ket{\psi}$ for a fixed and constant depth $D$. In practice, this may neglect symmetries that require exponentially long circuit depths to be implemented in the chosen PQC architecture.

\begin{figure*}[ht!]
    \centering
    \includegraphics[scale=0.45]{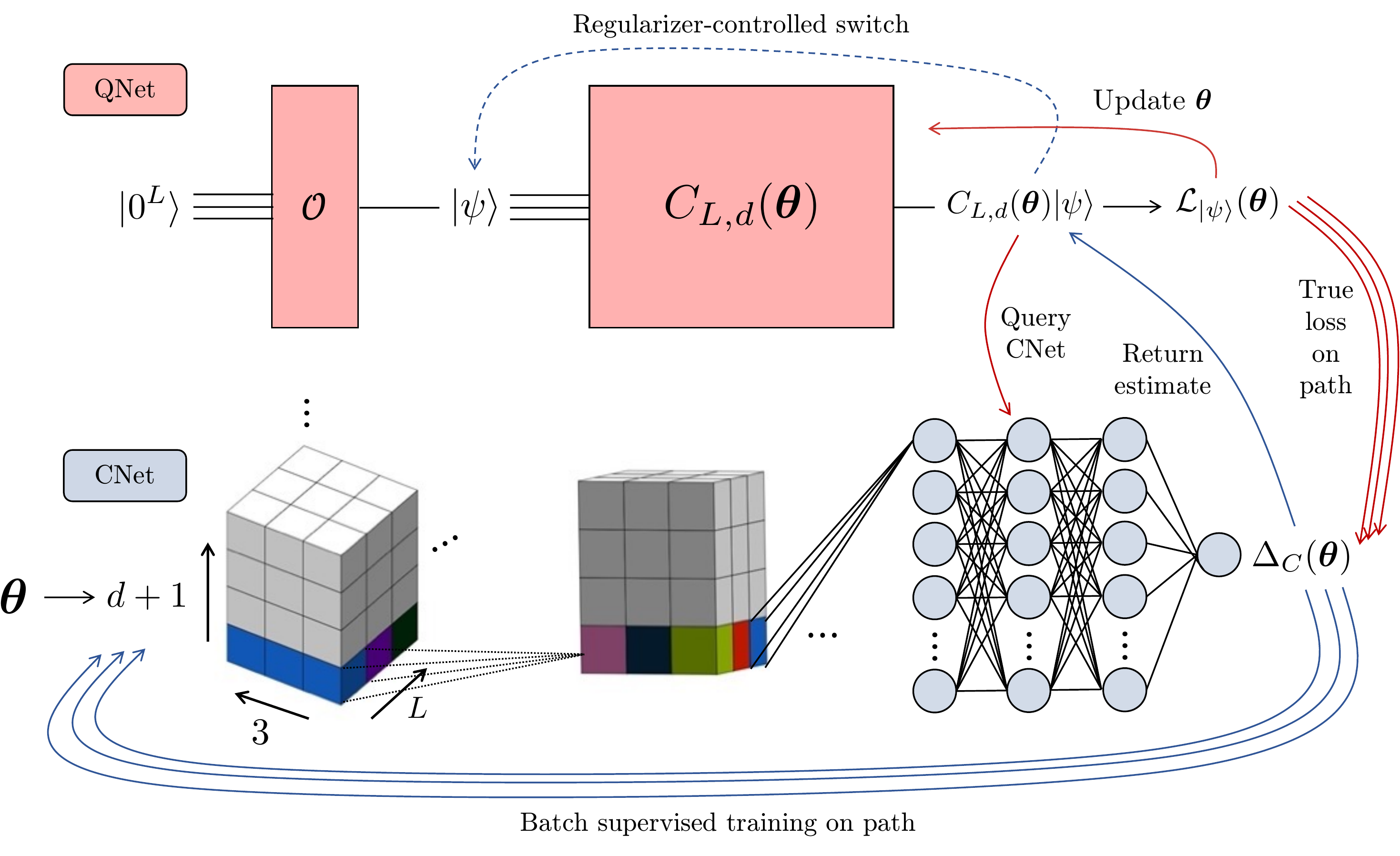}
\caption{Depiction of the symmetry learning algorithm. The quantum net (red) involves a variational quantum circuit $C_{L, d}(\bth)$ on $L$ qubits and with block-depth $d$ measured by a loss function $\CL_{\ket{\psi}}(\bth)$. $\bth$ is varied until $C_{L, d}(\bth)$ represents a symmetry. At the end of each epoch, a classical 3-dimensional convolutional deep net (CNet; blue) learns the loss function along the path just explored by the QNet. During future epochs, the CNet informs the QNet as to whether it has already explored its current path, and hence whether it needs to leave its current path to an as-yet explored region. The CNet structure matches the parameter structure of the QNet. The first layer convolves over each set of 3 parameters per qubit, and the second layer convolves over every parameter; the result is inputted into a 3-layer fully connected network.}
    \label{fig:learning_scheme}
\end{figure*}

We develop the symmetry learning scheme in three stages. First, we devise a method to verify whether a given operator $U$ is a symmetry. We then show that the verification procedure can be upgraded into a learning procedure, by introducing a variational quantum algorithm (VQA) built upon a PQC family $C_{L, D}$ of depth $D$ acting on $L$ qubits. Finally, we boost the learning scheme by introducing a regularization technique based on classical deep learning that prevents the VQA from repetitively proposing similar symmetries. The full scheme is illustrated in Fig.~\ref{fig:learning_scheme}, which we refer to throughout this section.

\subsection{\label{sec:theory:subsec:check}Symmetry Verification}
\begin{figure}[ht!]
    \centering


\includegraphics[scale=0.6]{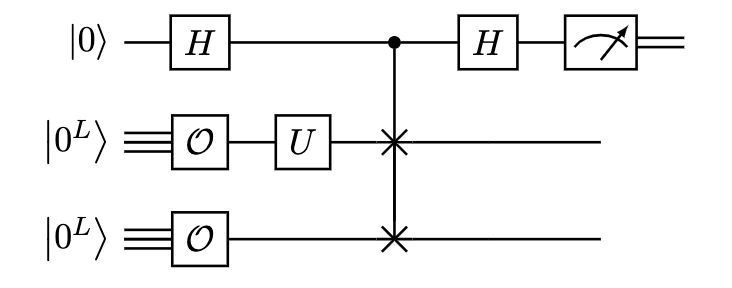}
    \caption{Verification circuit $V$ implementing a $\SWAP_\e$ test. The central operation is a swap of the two $L$-qubit registers (bottom registers), controlled by the ancillary qubit (top register). By running $V$ $O(1/\e^2)$ times, we can determine the overlap $|\bkop{\psi}{U}{\psi}|^2$ up to error $\e$. We estimate $U$ to be a symmetry of $\ket{\psi}$ if the overlap is at least $1 - O(\e)$.}
    \label{fig:swap_test}
\end{figure}

To begin, suppose that we are given an operator $U$ represented by a PQC family $C_{L, D}$. We restrict ourselves only to symmetries expressible in such a decomposition, noting that a universal quantum gate family can generally express any operator up to small error. Our first step will be to verify that $U$ is indeed a symmetry of $\ket{\psi}$. Such a problem can be decided in polynomial time by a method adapted from quantum fingerprinting, known as the $\SWAP_{\e}$ test~\cite{buhrman2001quantum}. The heart of the procedure is the controlled-SWAP gate, which takes $\ket{0} \leftrightarrow \ket{1}$ unitarily if a control qubit is $\ket{1}$. By utilization of the circuit in Fig.~\ref{fig:swap_test}, measurement of the ancillary qubit yields $\ket{0}$ with probability \begin{align}
    \Pr[\text{\texttt{0}}] = \frac{1}{2} + \frac{1}{2} |\bkop{\psi}{U}{\psi}|^2 .
\end{align}
Thus the overlap $|\bkop{\psi}{U}{\psi}|^2$ is a function of the bias (probability distance from $1/2$) of a coin flip, which can be determined to error $\e$ in $O(1/\e^2)$ trials~\cite{buhrman2001quantum}. Since the number of swaps scales linearly with qubit size, the time complexity of the $\operatorname{SWAP}_\e$ test is polynomial: $O(L/\e^2)$.

Although the $\operatorname{SWAP}_\e$ is efficient, it uses long-range SWAP gates spanning $O(L)$ qubits, which are not practical for general near-term quantum devices. In Section~\ref{sec:theory:sub:nisq}, we give an alternative verification procedure free of long-range couplings at the expense of computational efficiency.

\subsection{\label{sec:theory:subsec:quantum}Variational Quantum Generative Algorithm}
\begin{figure*}
    \centering
    \includegraphics[scale=0.6]{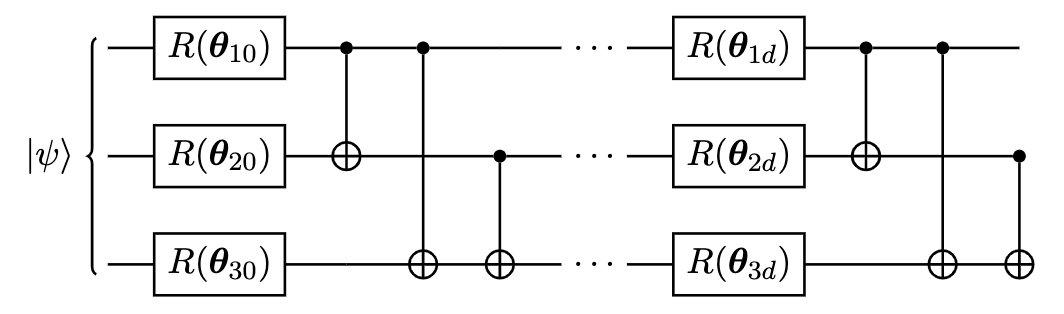}
    \caption{Example of a universal parameterized quantum circuit on $L = 3$ qubits. The cross-hairs are controlled-NOT gates and the $R$ gates are single-qubit rotations, parameterized by $\bth$. There are $d$ layers of $\binom{L}{2}$ controlled-NOT gates between each pair of qubits, sandwiched by the rotation gates.}
    \label{fig:universal_circuit}
\end{figure*}
In the process of symmetry verification we utilized the matrix element $|\bkop{\psi}{U}{\psi}|^2$ as a measure of the closeness $U$ is to a symmetry. The overlap may be interpreted as a loss metric, i.e. 
\begin{align}
    \CL_{\ket{\psi}}(U) = (1 - |\bkop{\psi}{U}{\psi}|^2)^2. 
\end{align}
This overlap can be estimated statistically by the $\SWAP_\e$ test. The formulation of verification as a loss allows for a transformation from verification---a discriminative procedure---to proposition (search)---a generative model. Consider again a universal PQC; one such family includes layers of $\binom{L}{2}$ CNOT gates that are sandwiched in between single-qubit rotation gates, as depicted in Fig.~\ref{fig:universal_circuit} for $L = 3$ qubits. This PQC ansatz is well-established and is known to be universal as the depth reaches exponential values~\cite{shende2003minimal}. In place of circuit depth, we use the \textit{block-depth} $d \geq 0$, the number of layers of $\binom{L}{2}$ CNOT gates. Hence, $d = 0$ corresponds to a single layer of rotations, $d = 1$ gives 2 layers of rotations sandwiching 1 layer of CNOT gates, etc. The circuit on $L$ qubits is specified uniquely by either the circuit depth $D$ or the block-depth $d$; we shall use the latter, writing the circuit as $C_{L, d}$ and filtering the symmetries by block-depth as \begin{align}
    S_0[\ket{\psi}] \seq \cdots \seq S_d[\ket{\psi}] \seq \cdots S_{O(\exp{L})}[\ket{\psi}] .
    \label{eq:sym_filter}
\end{align}
Under a paramterization $\bth$, let \begin{align}
    \CL_{\ket{\psi}}(\bth) = \CL_{\ket{\psi}}(C_{L, d}(\bth)) .
\end{align}
Our VQA generates symmetries as follows. Initialize the PQC randomly by drawing $\bth_0 \sim \Unif(0, 2\pi)^{3L(d+1)}$. Next, run $C_{L, d}$ on $\ket{\psi}$ and compute $\CL_{\ket{\psi}}(\bth_0)$. The loss will likely be large, but $\bth$ may be improved by an updating algorithm of classical machine learning. Our algorithm employs Nelder-Mead simplex search, which optimizes $\bth$ locally by updating the parameter to its simplex neighbor with the minimum loss value~\cite{nelder1965simplex}. The use of a classical updating scheme on a PQC has been adopted to considerable success in many hybrid quantum-classical algorithms such as variational quantum eigensolver (VQE) and QAOA. Here, $\bth$ is iteratively updated until $\CL_{\ket{\psi}}(\bth)$ falls below an error threshold $\d$, at which point the symmetry is proposed. See Appendix~\ref{app:hyperparam} for details on our choices of hyperparameters.

Since the PQC consists of layers of parameterized gates updated by a training algorithm, it has a natural interpretation as a quantum analog of a neural network, which we refer to as the QNet. The QNet VQA comprises the top (red) half of Fig.~\ref{fig:learning_scheme}.

\subsection{Symmetries as Manifolds}
The VQA proposed in the previous section above successfully generates symmetries; however, with no further assistance the PQC may propose the same symmetry repetitively. In the extreme case, the PQC could simply choose to always output the identity matrix. The prevention of repeat findings motivates the introduction of a regularization procedure. A na\"ive method might enumerate the set of symmetries found up to the $k$th epoch of search, $\bth_1, \hdots, \bth_k$, and then adding a regularization term to the loss function of the form \begin{align}
    \CL_{\ket{\psi}}^{(\text{reg})}(\bth) = \CL_{\ket{\psi}}(\bth) + \l \sum_{k'=0}^{k} \norm{\bth - \bth_{k'}}^{-2} . 
    \label{eq:crappy-reg}
\end{align}
Yet such regularizers are doomed to fail due to the algebraic structure of $S[\ket{\psi}]$. Not only are the products of symmetries also symmetries, generating $O(\exp{k})$ symmetries from a given set of size $O(k)$, but the symmetries are generally continuous families forming a Lie subgroup $S_d[\ket{\psi}] \leq \mathrm{U}(2^L)$. Consider, for example, the $L$-qubit GHZ state \begin{align}
    \ket{\mathrm{GHZ}_L} = \frac{\ket{0^L} + \ket{1^L}}{\sqrt{2}} .
    \label{eq:GHZ_state}
\end{align}
For simplicity, consider $d = 0$ (factorizable) symmetries $S_0[\ket{\mathrm{GHZ}_L}]$. Under a manifold picture of the Lie subgroup, the factorizable symmetries can be partitioned into two submanifolds \begin{align}
\begin{aligned}
    & \left\{ \bigotimes_{k=1}^L \twoMat{e^{i\a_k}}{0}{0}{e^{i\b_k}} \bigg\lvert \sum_{k=1}^L \a_k = \sum_{k=1}^L \b_k = 1 \right\} , \\
    &\left\{ \bigotimes_{k=1}^L \twoMat{0}{e^{i\a_k}}{e^{i\b_k}}{0} \bigg\lvert \sum_{k=1}^L \a_k = \sum_{k=1}^L \b_k = 1 \right\} .
\end{aligned}
\label{eq:GHZ-symmetries}
\end{align}
The first submanifold is composed of diagonal matrices with phases as their entries, and includes the identity matrix. The second submanifold is composed of off-diagonal matrices with phases as their entries, and includes the bit flip symmetry $X$ where $X$ is the Pauli spin operator $\sigma_x$. 
Each submanifold has four degrees of freedom, forming a four-torus $T^4$. These may be equivalently expressed in circuit decomposition as \begin{align}
\begin{aligned}
    M_1 & = \left\{ \bigotimes_{k=1}^{L} X P(\a_k) X P(\b_k) \right\} , \\
    M_2 & = \left\{ \bigotimes_{k=1}^{L} P(\b_k) X P(\a_k) \right\} ,
\end{aligned}
\label{eq:GHZ-Mi}
\end{align}
where $P(\varphi)$ is the phase gate \begin{align}
    P(\varphi) = \twoMat{1}{0}{0}{e^{i\varphi}}.
\end{align} 
Each submanifold $M_i$ is a smoothly parameterized set of symmetries. Thus, in principle we need only inspect a few points in the submanifold to deduce a closed-form expression for $M_i$. 

\subsection{\label{sec:theory:subsec:regg}Interactive Classical Learning}

To learn efficiently, we cannot explore the same submanifold repeatedly while other manifolds remain unexplored. Rather, we must develop a procedure to push the PQC away from symmetry submanifolds already explored sufficiently.
While such a task is not obvious, it is not impossible from a measure-theoretic standpoint. That is, there will never be a case in which $\ket{\psi}$ has two symmetry submanifolds $M_1, M_2$ such that $\dim(M_1) < \dim(M_2)$, so that $M_1$ would never be found. For as the manifolds are also groups, there exists automorphisms mapping $M_1$ to $M_2$ and vice versa, and hence the topological dimension of every symmetry submanifold must be identical.
Nonetheless, no pointwise regularization procedure, such as that of Eq.~(\ref{eq:crappy-reg}), can recognize two symmetries being of the same submanifold and thereby push the PQC away from proposing both of them.

We therefore introduce a regularization method based on classical deep learning: rather than keeping track of specific symmetries, we train the classical net (CNet) to approximate $\CL_{\ket{\psi}}(\bth)$ by its output $\hat{\CL}_{\ket{\psi}}(\bth)$. The direct statistical estimation of $\CL_{\ket{\psi}}(\bth)$ requires quantum operations and measurement handled by the QNet, so we supervise the CNet using the QNet itself. Specifically, for every batch of loss-functional evaluations the QNet performs as it searches for symmetries, we train the CNet in parallel using those evaluations as the supervisory dataset. Each batch is a transcript of the path in $\mathrm{U}(2^L)$ (henceforth abbreviated to $\mathrm{U}$) walked by the QNet that terminates at a symmetry. Ultimately, the aim of the CNet is to estimate with low error the loss for $\bth$ in a neighborhood of each $\bth$ explored by the QNet. That is, we train the CNet by minimizing the squared error \begin{align}
    \D_C(\bth) = (\CL_{\ket{\psi}}(\bth) - \hat{\CL}_{\ket{\psi}}(\bth))^2
    \label{eq:MSE}
\end{align}
where $\hat{\CL}_{\ket{\psi}}$ is the CNet estimate of the true loss.

Since the landscape of the loss function can be highly complex, we cannot hope to obtain a good classical estimate of the loss globally with only a few local training points. Instead, we design the CNet to estimate local patches around points (neighborhoods of the point in $\mathrm{U}$) explored by the QNet with very low error, but estimate patches far from points explored by the QNet with high error. Equivalently, the CNet should overfit purposefully. Thus, by analyzing the CNet estimation error on a point $\bth$, the QNet can determine whether $C_{L, d}(\bth)$ resembles a symmetry previously explored (i.e. the absolute difference of their overlaps are small). If so, the QNet can adjust accordingly by leaving the local region of parameter space to hopefully find another region of $\mathrm{U}$ that has not yet been explored. We found numerically (see Appendix~\ref{app:cnet}) that the average difference in $\D_C$ evaluated on a training point versus a random point is about two orders of magnitude, making for easy distinction.

Classical deep neural nets are well-suited to the task because a CNet can extrapolate relatively well locally. Consequently, as we will also show empirically, the CNet accomplishes what no pointwise regularization scheme can: it regularizes against entire continuous families of symmetries. Since we train the CNet on the entire search path, including points in $\mathrm{U}$ not corresponding to symmetries, the CNet regularizes against any subsets of $\mathrm{U}$ that the QNet has already explored, even if they are not symmetry submanifolds. Thus, the CNet minimizes the time the QNet wastes exploring parts of $\mathrm{U}$ already well-understood. Eventually, as the QNet explores more and more symmetries, all of the symmetry submanifolds of fixed $d$ will be learned. The CNet will then find low estimation error everywhere, at which point the learning algorithm will terminate. Thus when classical learning is added to the QNet VQA, we find empirically that the symmetry learning process converges.

As shown in the bottom half of Figure~\ref{fig:learning_scheme}, we develop a CNet with two three-dimensional convolutional layers. We design the convolutions to respect the structure of the parameterization of the PQC. In particular, the first layer scans over each set of $3$ rotation parameters per qubit per block-depth, and the second layer scans over each individual parameter. The convolutional net layer is connected to a simple three-layer fully-connected neural net with leaky rectified linear unit (ReLU) activation functions~\cite{agarap2018deep}.

We train the CNet in path-batches; during each symmetry-learning epoch, the CNet reports $\D_C$ while the QNet walks Nelder-Mead paths in parameter space, corresponding to a path in $\mathrm{U}$. At the end of each epoch, when a symmetry is proposed, we use batch stochastic gradient descent (SGD) on the transcript of points walked in the path.
Note that the CNet cannot undergo training during an epoch of QNet symmetry search. Otherwise, the CNet will extrapolate the local path well and mislead the QNet into thinking that it has already traversed its current path. We elect to use SGD instead of an algorithm similar to Nelder-Mead for its superior empirical performance (see Appendix~\ref{app:cnet}).

At Every $N$ steps on the QNet search path, the QNet pauses the search and queries the CNet. If the CNet estimation error is low, then the QNet must adjust its search path accordingly. We propose two methods by which the QNet can redirect its search. In the first, which we call the global method, the QNet simply randomly restarts. Since the parameter dimension is \begin{align}
    \dim \bth = O(L d) ,
\end{align}
the dimensionality of search space is sufficiently large for even moderate $L, d$ that random restarts give a substantial likelihood of landing in an as-yet traversed region of $\mathrm{U}$. $N$ is chosen to be much smaller than the number of steps to find a symmetry, but still reasonably large ($\sim 100$), in consideration of a tradeoff between wasting too much time querying the CNet and wasting too much time searching a familiar region.\par
The second method, which we call the local method, alternates between finding symmetries (optimization) and maximizing the classical estimation error $\D_C$ (exploration). That is, instead of randomly jumping to a new region, we directly walk in the direction that appears most unfamiliar. In the exploratory phase, we use finite difference gradient descent implemented by the shift rule~\cite{grossmann2007numerical}, 
\begin{widetext}
\begin{equation}
    \bth \leftarrow \bth - \eta \br{\frac{\CL_{\ket{\psi}}(\th_{101} + \d, \th_{102}, \hdots, \th_{3d3}) - \CL_{\ket{\psi}}(\th_{101} - \d, \th_{102}, \hdots, \th_{3d3})}{2\d}, \cdots} ,
\end{equation}
\end{widetext}
where $\eta$ is the learning rate and $\d$ is the finite difference parameter. We favor the shift rule over Nelder-Mead because the latter fails to update $\bth$ if all points in the neighboring simplex have similar $\D_C$; hence, no exploration occurs. By contrast, gradient descent will still update $\bth$, allowing progressive movements away from known regions. The local method is more efficient for states that are highly symmetric (i.e. many symmetry submanifolds close to each other, for a reasonable metric of submanifolds on $\mathrm{U}$). The global regularizer has an advantage on states with high-dimensional parameter spaces, wherein random restarts are generally faster than explicitly carving paths to unexplored regions.

The full learning scheme is given in Figure~\ref{fig:learning_scheme}. As Nelder-Mead has been shown to be average-case efficient, under the assumption that the regularizer prevents the same manifold from being explored more than polynomially many times, our algorithm solves the symmetry learning problem efficiently. In the absence of regularization, symmetry learning may take superexponential time or even fail to terminate at all.

\subsection{ \label{sec:theory:sub:nisq}Adjustments for Near-term Practical Implementation}
Due to the limitations of near-term quantum circuits, the symmetry learning algorithm as formulated above will be difficult to implement with current hardware. Specifically, both the $\operatorname{SWAP}_\e$ test and the PQC given in Fig.~\ref{fig:universal_circuit} are spatially nonlocal computations---they contain gates that span $O(L)$ qubits. In this section, we adjust the algorithm to use only local operations, thereby becoming realizable with near-term hardware, at the cost of some efficiency and universality.

\begin{figure}[ht!]
    \centering
    \includegraphics[scale=0.6]{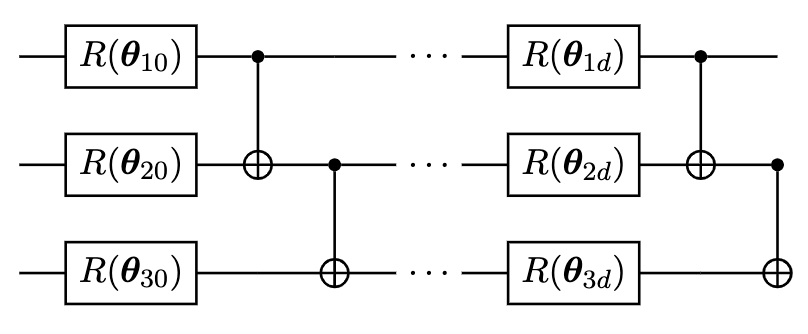}
    \caption{Restricted quantum circuit family, of $d$ layers, appropriate for current noisy devices capable of local operations. Cross-hairs represent controlled-NOT gates, while $R$ gates are single-qubit rotations, parameterized by $\bth$.}
    \label{fig:PQC}
\end{figure}

A SWAP operation between two $L$-qubit registers is implemented via $L$ single-qubit SWAP gates spanning $O(L)$ qubits. While recent advancement in Rydberg atom quantum computers have made long-range SWAPs a reality~\cite{bluvstein2021quantum}, in more general quantum hardware the SWAP gate must be decomposed into controlled-NOT gates as \begin{align}
    \SWAP_{ij} = \CNOT_{i\to j} \CNOT_{j\to i} \CNOT_{i \to j} .
    \label{eq:swap_cnot}
\end{align} 
Hence a controlled-SWAP can be carried out in $O(L)$ Toffoli gates. However, such gates span $O(L)$ qubits. An alternative verification procedure to the $\operatorname{SWAP}_\e$ test is via statistical comparison of the projective measurements (i.e. their distribution in a particular measurement basis) of $\ket{\psi}$ with that of $U \ket{\psi}$.
For any fixed basis $b$, we may estimate the classical Kullback-Leibler (KL) divergence $\KL_b(\ket{\psi}, U\ket{\psi})$ between the measurement distributions of $\ket{\psi}$ and $U \ket{\psi}$ by constructing a sampling-based empirical cumulative distribution function (ECDF) for each state in the basis expansion of $b$.
Choosing multiple bases to capture the structure of the phases, we define the quantum KL divergence as a sum over the KL divergence of each projective measurement: \begin{align}
    \QKL_{\ket{\psi}}(U) = \sum_{b} \KL_b(\ket{\psi}, U\ket{\psi}) .
\end{align}
In practice, we find that only two bases, which need not be random, are required to capture the contributions of the phases (see Appendix~\ref{app:2bases}). Generally, estimation of $\QKL_{\ket{\psi}}(U)$ requires $O(\exp{L})$ evaluations of $U$ just to sample every basis element at least once on average. However, if only a small (polynomial) subset of the basis elements have non-negligible amplitude, the QKL loss may be evaluated efficiently. Thus, although the exponential search is generally unavoidable, we can first search for a polynomially sparse basis by checking the $z$ and $x$ bases and choosing the sparser one. Further optimizations can be done via a method of binary search to find a sparse basis. A second basis can then be found by rotating the sparsest basis by a small angle (we chose $\pi/10$). Such basis optimizations should be done before beginning the learning  algorithm, since checking a basis requires one sampling round, whereas the learning algorithm will use hundreds of thousands or more sampling rounds, one for each step in the search path. With preliminary basis finding, the QKL verification remains relatively efficient, and we define the near-term loss as \begin{align}
    \CL_{\ket{\psi}}(\bth) \stackrel{\text{near-term}}{\longrightarrow} \QKL_{\ket{\psi}}(C_{L, d}(\bth)) .
\end{align}
The second modification is of the PQC ansatz. The form given in Fig.~\ref{fig:universal_circuit} is universal, but has $\binom{L}{2}$ CNOT gates in each layer, most of which span $O(L)$ qubits. Ideally, a hardware-efficient circuit family has only gates spanning $O(1)$ qubits and scales in depth at most linearly with $L$. Thus, we restrict the CNOT gates to only nearest neighbors, giving a hardware efficient circuit shown in Fig.~\ref{fig:PQC}. Although the family may no longer be universal, or at least require much longer depths to represent a general unitary, it can still represent most symmetries of practical interest, including any expressible in terms of local couplings.
For present applicability, we adopt both adjustments in this paper. Note that our algorithm is agnostic to the choice of PQC family and loss function, so these hardware efficient modifications can be readily removed when the appropriate experimental hardware is realized.

\section{\label{sec:results}Results and Discussion}
We demonstrate our results on four quantum states: a Bell state, the GHZ state from Eq.~(\ref{eq:GHZ_state}) and the planar cluster state~\cite{Gottesman1999}, which both arise from a broader family known as stabilizer states, and the transverse Field Ising Model (TFIM) which is known as one the simplest quantum spin chains.
Much of our focus is on the GHZ state, for due to their sensitivity to decoherence, they are often utilized for characterization of noisy quantum hardware \cite{omran2019generation,Song2019,Wei2019} and other quantum technologies, such as error correction, quantum metrology, and quantum communication \cite{Zhao2004,Pezze2018}.
We will first visualize some symmetries of the states. We then benchmark the scaling of the algorithm with respect to $L$ and $d$ as well as its robustness against noise. We collect data by backend classical simulation in Qiskit~\cite{Qiskit}, although our results on simulated noise indicate comparable results on quantum hardware.

\subsection{Symmetries of Simple Circuit States}

\begin{figure}[ht!]
    \centering
    \includegraphics[scale=0.5]{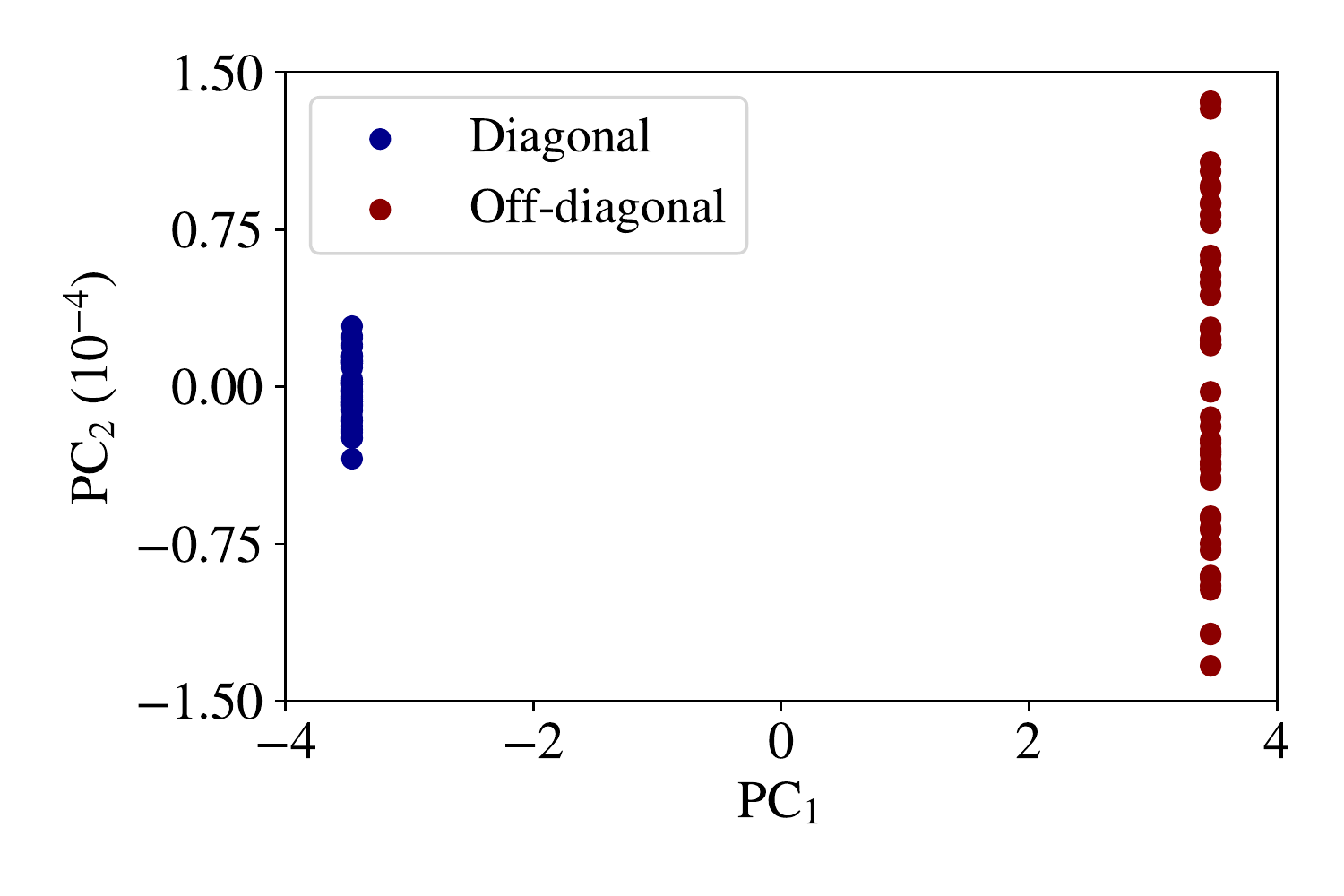}
    \caption{PCA projection of 100 symmetries of the 3-GHZ state, found by the QNet, into 2 dimensions. The separation between them implies that the parameterizations of the manifolds are easily distinguishable, which we know analytically to be true.}
    \label{fig:PCA}
\end{figure}

The exponential size of the matrix representation of $\bth$ implies that there is no simple method to visualize a general symmetry aside from its circuit decomposition; we discuss further this issue at the end of the paper. However, for simple cases, we can explicitly examine the geometric or analytical structure of the symmetry, which we discuss for the $L$-qubit GHZ state, measured in the $z$ basis for maximal sparseness. When $d = 0$, each symmetry factors as a tensor product of $2 \times 2$ matrices, which belong to one of the submanifolds in Eq.~(\ref{eq:GHZ-Mi}). Since they are either diagonal or off-diagonal, we use principal component analysis (PCA) in 2 dimensions to visualize their geometric structure, illustrated in Fig~\ref{fig:PCA} for $L = 3$. The diagonal and off-diagonal symmetries form distinct clusters, which is consistent with the orthogonality of their analytical representations.

Beyond $d = 0$, no such tensor product decomposition exists, so the symmetries become more difficult to visualize. (One can do so, though, by using PCA on $\bth$ itself and then clustering the resultant dataset.) However, for $L = 2$, the unitary implemented by $C_{2, d}$ may be explicitly written. A key symmetry in many entangled states is the swap symmetry, which require at least $d = 3$ to be represented, by Eq.~(\ref{eq:swap_cnot}). We test the algorithm on a Bell basis state \begin{align}
    \ket{\Phi} = \frac{\ket{01} + \ket{10}}{\sqrt{2}}
\end{align}
to numerically obtain the swap symmetry. On average, after $\sim 10$ trials, we find many symmetries in the family \begin{align}
U = e^{i\varphi} \begin{pmatrix}
\cdot & c_1 & -c_1 & \cdot \\
\cdot & \a & 1-\a & \cdot \\
\cdot & 1-\b & \b & \cdot \\
\cdot & c_2 & -c_2 & \cdot 
\end{pmatrix} ,
\label{eq:swap_matrix}
\end{align}
where $\a, \b, c_j \in \C$ and the dots representing arbitrary numbers to make $U$ unitary. In the limit of $c_j, \a, \b \to 0$, $U$ reduces to the familiar SWAP operation, but the algorithm finds a more general family that may be interpreted as a partial swap operation on each of the $\ket{01}$ and $\ket{10}$ basis states. More generally, we find a similar family for the $L$-GHZ state, with a copy of the $2 \times 2$ submatrix at the center of Eq.~(\ref{eq:swap_matrix}) present in each two-qubit subspace. This is a general principle of symmetry learning: one samples from a smooth symmetry submanifold, and must take a few samples to interpret the underlying family of symmetries being represented.

\subsection{Symmetries of Adiabatically Prepared States}

\begin{figure*}[ht!]
    \centering
    \includegraphics[scale=0.5]{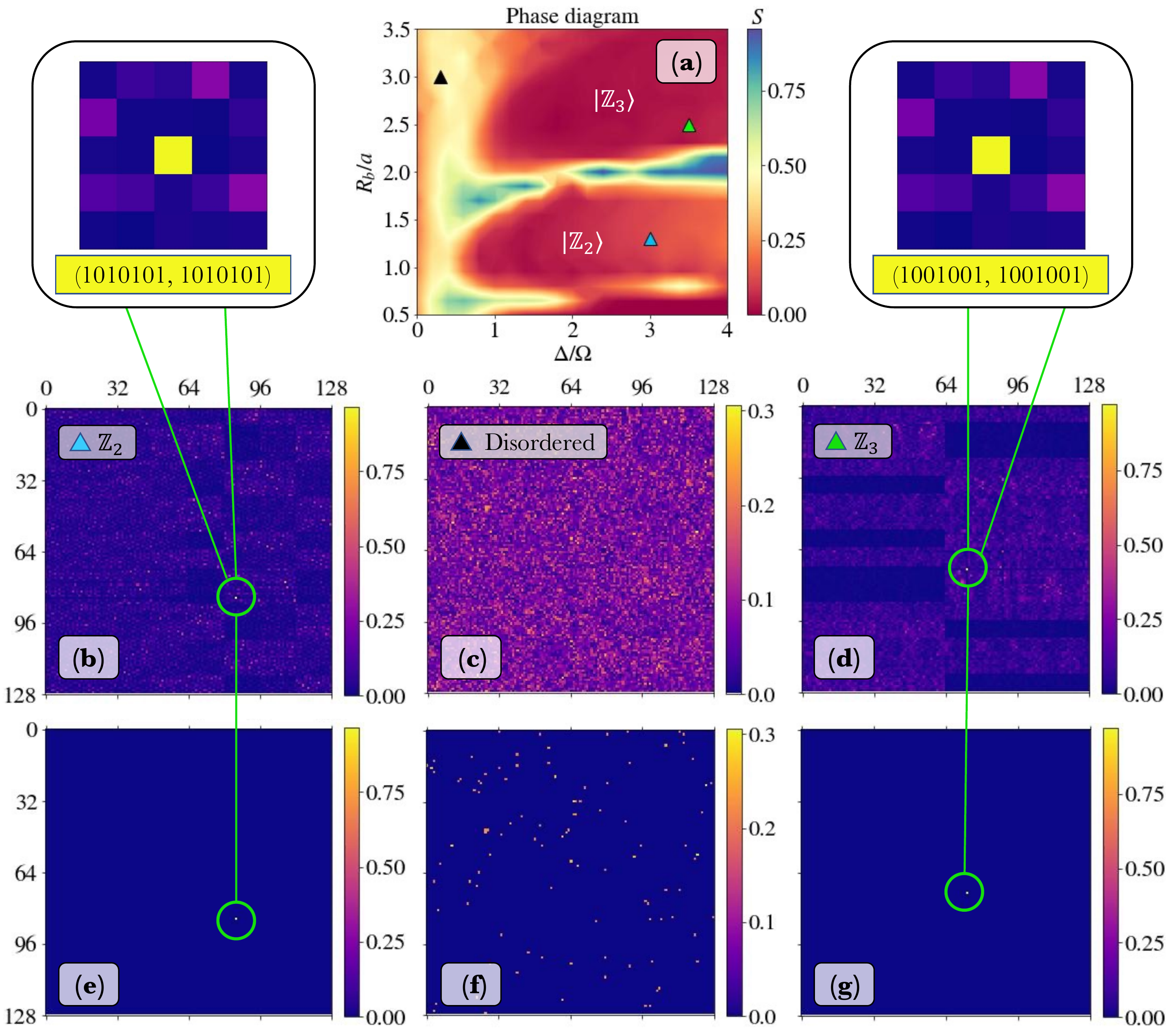}
    \caption{\updated{Visualization of the symmetries of the 7-qubit Rydberg chain $\ket{\D/\W, R_b/a}$, whose ground state is prepared by adiabatic evolution. (a): the phase diagram based on the entanglement entropy $S$ of $\ket{\D/\W, R_b/a}$ with two ordered lobes, (b)-(d): matrix visualizations $|U_{ij}|^2$ of a representative symmetry $U \in \R^{2^7 \times 2^7}$ for three Rydberg ground states. Most of the matrix elements look like random noise, but on the $\Z_2$ and $\Z_3$ there is a clear single element that is close to 1, which is zoomed in on the inset. The ``signal" matrix elements are more easily seen in (e)-(g), which send all elements where at least $\a = 0.1$ smaller than the maximum element to 0.}}
    \label{fig:Rydberg_analysis}
\end{figure*}

\updated{
In the following section, we will investigate our algorithm on further families of states. Before doing so, we note that we have yet to use the black-box assumption (that no information is known about the state) of the states. The motivation for black-box states is that in experiment one may not have efficient access to the inverse circuit used to prepare the state because the preparation may involve, for example, unitary time evolution. This is clearly different for the GHZ family. Often states that require time evolution are both the most interesting and the most difficult to study analytically. As such, learning their symmetries sheds insight into the state itself and provides utility of our algorithm beyond the computational boosts discussed in Section~\ref{sec:intro}. One important example of a family of states prepared with time evolution is the Rydberg family. Specifically, consider the 1-dimensional Rydberg atom chain defined as the ground state of the Hamiltonian
}
\begin{align}
\begin{aligned}
    \frac{\mathcal{H}(t)}{\hbar} & = \sum_j \frac{\Omega_j(t)}{2}(e^{i \phi_j(t)}|0_j\rangle\langle 1_j|+e^{-i \phi_j(t)}| 1_j\rangle\langle 0_j|) \\
    & \phantom{hii} -\sum_j \Delta_j(t) \hat{n}_j+\sum_{j<k} V_{j k} \hat{n}_j \hat{n}_k,
\end{aligned}
\end{align}
\updated{
where $\hat{n}_j = \ketbra{1_j}{1_j}$, $\W_j$ is the Rabi frequency for qubit $j$, $\p_j$ is the laser phase, and $\D_j$ is the laser detuning frequency. $V_{jk}$ is the potential, which follows the van der Waals power law $V_{jk} = C_6 / |\mb{r}_i - \mb{r}_j|^6$, where $C_6 = 862 \times 2 \pi$ MHz $\m$m$^6$. In recent years interest in Rydberg atoms has exploded due to their potential as scalable neutral-atom quantum computers and programmable simulators, as well as their interesting quantum dynamics~\cite{ebadi2022quantum,adiabatic-1D,adiabatic2D,gomez2022reconstructing,levine2019parallel,keesling2019quantum}. When the parameters are time-independent and uniform, i.e., $\W_i = \W_j$ and $\D_i = \D_j$ for all $i$ and $j$, the (matter) phase of a Rydberg atom array is determined by two parameters, $\D/\W$ and $R_b/a$, where $a$ is the lattice constant and $R_b$ is the blockade radius defined by $C_6/R_b^6 = \W$. We thus label states as $\ket{\D/\W, R_b/a}$. Physically, the blockade radius is the minimum distance between two Rydberg atoms for them to be simultaneously excitable, and is closely related to the NP-complete problem of Maximum Independent Set~\cite{ebadi2022quantum,pichler2018quantum}.
}

\updated{
For concreteness, we study the 7-qubit Rydberg chain with open boundary conditions, setting $\p_j = 0$. Under these conditions, the Rydberg chain has two ordered phases, denoted $\Z_2$ and $\Z_3$, which are nearly classical because they look like $\ket{\Z_2} = (1-\e)\ket{1010101} + \e\ket{\text{other}}$ and $\ket{\Z_3} = (1-\e) \ket{1001001} + \e \ket{\text{other}}$. In the literature, these phases are also said to obey a discrete permutation symmetry because the $\Z_n$ phase is invariant, up to a small error, when cyclically permuted $n$ times~\footnote{This definition, while useful in our analysis, is imprecise in general. A more accurate formulation of ordered phases is through the construction of an order parameter that takes on a certain range of values in each phase, with sharp changes at the boundary. While the order parameter unambiguously defines the ordered phase for any system size, the closeness-to-a product-state property does not hold for large systems because of quantum fluctuations. As each qubit in the chain has a small error, the total error from a product state scales exponentially with system size, and has been observed in experiment by~\citet{adiabatic2D}. We study this example as a simple system---small enough to be classically simulated---with a physically meaningful and important symmetry.}. Symmetry learning provides an alternative way to probe ordered phases numerically. Although there are much easier ways to deduce this symmetry, such as just ansatzing some possible spatial translation symmetries and checking them explicitly, this example provides evidence that symmetry learning can become useful to learn much more complicated symmetries of physically interesting ground states in the future.
}

\updated{
To simulate the Rydberg arrays, we use the technique of adiabatic evolution. Adiabatic evolution achieves the ground state of a target Hamiltonian by starting with an easily preparable ground state of an initial Hamiltonian, and then slowly modifying the Hamiltonian to reach the target. In practice, the time scale of the parameter variation is $\sim 3 \,\m$s. This variation can be thought of as iteratively time-evolving the state according to a smoothly parameterized Hamiltonian, resulting in the state \begin{align}
    \ket{\psi(t)} = e^{-i \int dt \, \CH(t)} \ket{\psi(0)} .
\end{align} 
Adiabatic evolution is a key technique to Rydberg-atom quantum computing~\cite{adiabatic-1D,samajdar2020complex}. We set $\ket{\psi(0)} = \ket{0}$ for simplicity, though any easily preparable state works equivalently.
We prepare Rydberg states in classical simulation using Hamiltonians generated from \texttt{Bloqade}~\cite{bloqade}. Having simulated experimental conditions, we proceed to examine their symmetries. Figure~\ref{fig:Rydberg_analysis}(a) shows the phase diagram of the 7-qubit Rydberg chain, calculated by the entanglement entropy \begin{align}
    S[\ket{\D/\W, R_b/a}] = -\operatorname{Tr} [\r_A \log \r_A]
\end{align}
where $\r_A$ is the reduced density matrix of $\ket{\D/\W, R_b/a}$ calculated by tracing over half of the chain. The two lobes correspond to the ordered phases. From the diagram we examine three points: $\ket{3.0, 1.3}$ in the $\Z_2$ phase, $\ket{3.5, 2.5}$ in the $\Z_3$ phase, and $\ket{0.3, 3.0}$ in neither (the disordered phase). Figures~\ref{fig:Rydberg_analysis}(b)-(d) respectively visualize a representative symmetry from these phases, by plotting $|U_{ij}|^2$ for each.\footnote{Our algorithm found these symmetries and optimized to losses of 0.062 for $\Z_2$, 0.0428 for $\Z_3$, and 1.416 for disordered, over 2000 iterations. Further details are in the Appendix.} Similar results are found by repeatedly running the algorithm until the CNet stops it. In (b) and (d), almost all elements look like random noise except for a single bright point, which we zoom in on in the insets. In binary, the indices correspond to exactly the $\Z_2$ and $\Z_3$ phases, since they are respectively close to $\ket{1010101}$ and $\ket{1001001}$. As such, the corresponding matrix element in the symmetries are close to 1, while all others are random noise far below 1. This visual is even more clear when we choose a cutoff $\a$ and set each element $|U_{ij}|^2$ to 0 unless $|U_{ij}|^2 \geq \max_{ij} |U_{ij}|^2 - \a$; this is shown in (e)-(g) for $\a = 0.1$. On the other hand, the ``symmetry" for the disordered phase, which has no special structure, resembles random noise, both in (c) and (f). This lack of structure reflects the disordered nature of the state. 
}

\updated{
    Three observations are of importance in the Rydberg example. First, although we used adiabatic preparation which produced a much more imperfect phase diagram as compared to exact diagonalization (which has smooth and almost entropyless ordered phase lobes), we nonetheless managed to learn symmetries. This illustrates the robustness of our algorithm to the more noisy and imperfect experimental settings and will be explored further in the next section via precise noise models. Second, we deduced from Fig.~\ref{fig:Rydberg_analysis} that the symmetries of the $\Z_2$ phase were matrices $U$ of the form $U_{85, 85} = 1 - \a_{85,85}$ ($85 = 1010101$ in binary) and all other $U_{ij} = \a_{ij}$, where the $\a$'s are small. This large ``signal" matrix element indicates that the state itself is nearly classical, hinting that the phase is highly ordered. The same analysis follows for the $\Z_3$ phase, and the opposite for the disordered phase, which allow us to conclude that there is no simple structure for disordered states. Thus, from symmetry learning, we have deduced most of the salient aspects of the Rydberg phase diagram, which in turn shed light on the physics. Symmetry learning, therefore, provides a new way to probe physics described by Hamiltonians. Lastly, while Fig.~\ref{fig:Rydberg_analysis} shows that there is no simple structure to the symmetries of the disordered phase, little further insight can be obtained by inspecting the matrix alone. We discuss this problem further in Section~\ref{sec:conclusion}.
}

\updated{
As we conduct numerical scaling analyses in the next section, we continue to consider ground states of Hamiltonians that require the black-box assumption, but turn to an Ising model for simplicity.
}

\subsection{Larger Degrees of Freedom}
As shown in Eq.~(\ref{eq:sym_filter}), more symmetries are representable by $C_{L, d}$ with increasing $d$. Inversely, the minimum $d$ required for the learning algorithm to find symmetries is a state complexity measure. For example, although every $S_0[\ket{\psi}]$ is nonempty for any $\ket{\psi}$ (due to the identity matrix), the rotation symmetries among all possible single-qubit rotation operations may be sparse and thus difficult to find. Thus, relatively few epochs will successfully find a symmetry. For a given $d$, the average loss over many (say, 100) epochs on a given state measures the difficulty of learning symmetries at that block-depth. (We assume implicitly here that trainability is not a problem, which breaks down for $d$ larger than what we probe here, due to the general vanishing gradient problem in machine learning.) We can examine the average loss on various states to compare the complexity of their symmetries---that is, how large $d$ must be before most epochs can find symmetries. Importantly, this simple measure can be done without knowing the actual symmetries themselves; we do not even need to look at the symmetries to understand the relation between the loss and the block-depth.\par

\begin{figure}[ht!]
    \centering
    \includegraphics[scale=0.45]{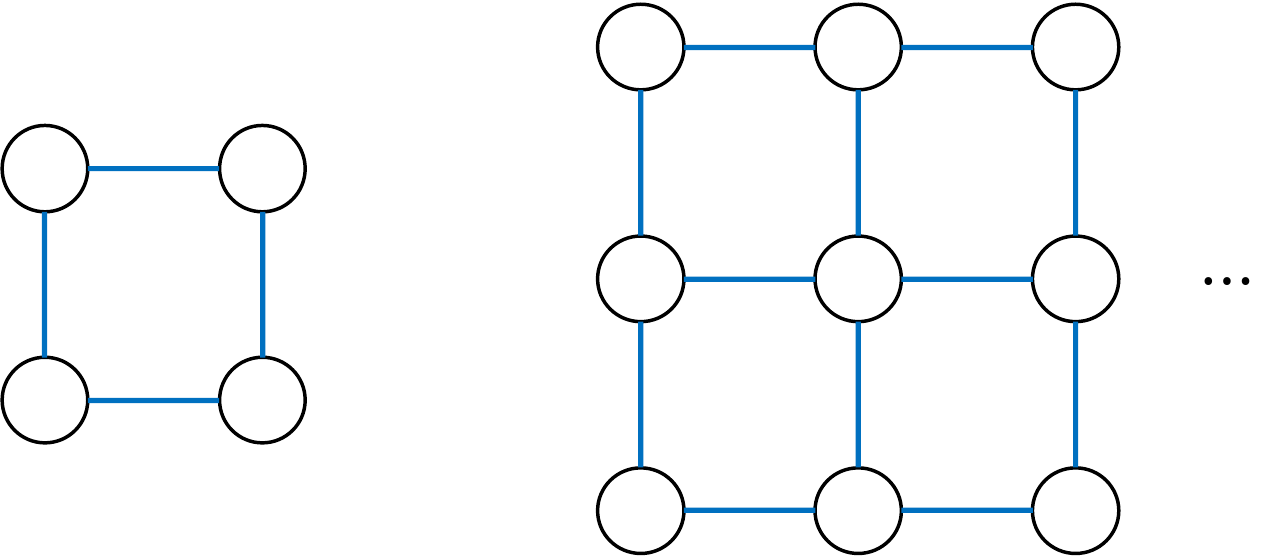}
    \caption{Graph representation of planar cluster states for $L = \ell^2$ qubits. The nodes are qubits and the edges are controlled-$Z$ gates. The cluster state is thus defined as $\Pi_{(a,b)\in \text{Edges}}C_{a}Z_b|+\rangle^{\otimes L}$ where $C_{a}Z_b$ is a control-$Z$ gate between qubits $a.b$ and  $|+\rangle=\frac{1}{\sqrt{2}}(|0\rangle+|1\rangle)$.}
    \label{fig:app:cluster_state}
\end{figure}

We demonstrate the above procedure on the GHZ state, the planar cluster state, and the ground state of the TFIM Hamiltonian. The planar cluster state is represented as a graph state for which nodes are qubits and edges are controlled-$Z$ gates, where $Z$ is the Pauli $\s_z$ operator. The planar cluster states are defined only for $L = \ell^2$ qubits, and are shown in Fig.~\ref{fig:app:cluster_state}.
The TFIM Hamiltonian is given by 
\begin{equation}
    \CH = 0.5 \sum_{i} S_{i}^{z} - J\sum_{i} S_{i}^{x} S_{i+1}^{x} ,
\end{equation}
where $J = 1$ is the interaction strength and the $\sigma_i$ are the Pauli spin operators on the $i$th qubit. While such variational methods as VQE and QAOA can be used to prepare the TFIM ground state, we do so numerically via exact diagonalization.

\begin{figure}[ht!]
    \centering
    \includegraphics[scale=0.5]{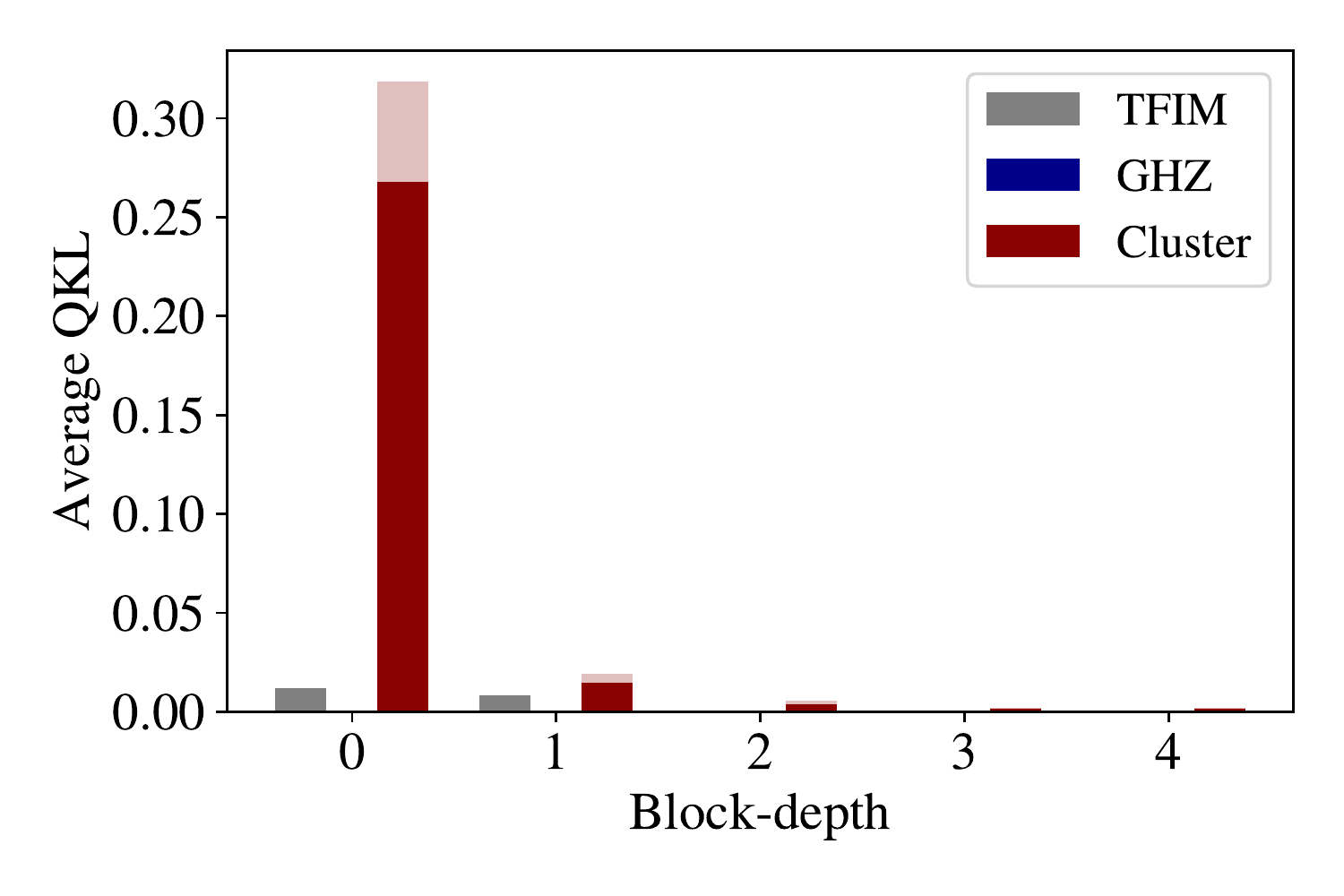}
    \caption{Average loss for 4-qubit GHZ, TFIM, and cluster states as a function of $d$. The GHZ losses are all negligibly small and thus invisible. The transparent bar indicates one standard deviation.}
    \label{fig:CC}
\end{figure}

Our preliminary analysis shows that the TFIM state is maximally sparse in the $x$ basis and cluster/GHZ states in the $z$ basis. Finding 100 symmetries for each depth with $L = 4$ yields an average loss given in Fig.~\ref{fig:CC}.
As we might expect, the cluster state has the most complex symmetries, requiring $d \approx 2$ to easily find symmetries with every epoch. The GHZ state is the simplest due to its well-studied $d = 0$ symmetries from Fig.~\ref{fig:PCA}.
In practice, we can resolve the issue of hard-to-find symmetries for small $d$ by postselecting on epochs that do find symmetries, using regularization to prevent traversing known paths. However, Fig.~\ref{fig:CC} shows that for relatively small $d$ even complicated states like the cluster state permit most epochs to learn symmetries, evidencing that our hardware-efficient PQC ansatz remains sufficiently rich to efficiently parameterize non-trivial symmetries.

\subsection{Query complexity}
We benchmark the numerical scalability of the learning algorithm both with respect to $d$ and $L$, using the $L$-GHZ state as a concrete example. We measure scalability in terms of the search query complexity; that is, the average number of Nelder-Mead iterations per epoch of symmetry learning. The advantage of query complexity is that it is agnostic to the verification procedure, so our results hold even without the near-term related modifications.
\begin{figure*}[ht!]
    \centering
    \includegraphics[scale=0.5]{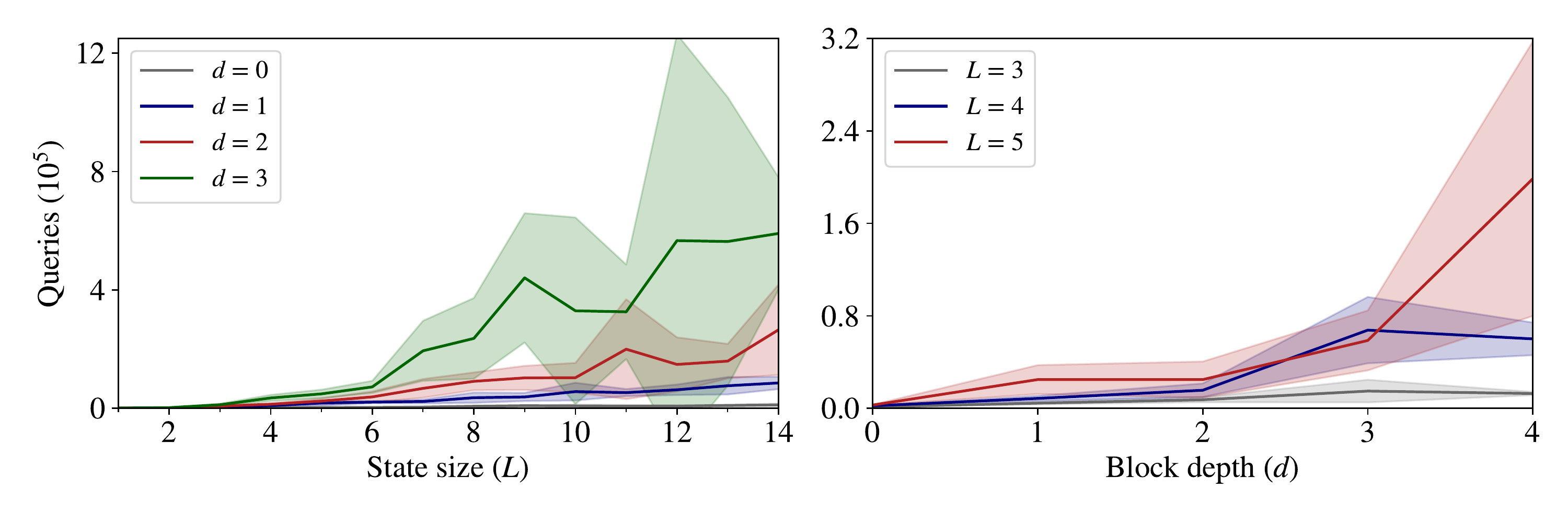}
    \caption{Average query complexity (number of search iterations to find a symmetry) of the GHZ state as a function of state size $L$ (left) and block-depth $d$ (right), estimated with $10$ trials per $L$ per $d$. Shading represents standard deviation.}
    \label{fig:DS}
\end{figure*}
With current devices, short-depth implementations are of the most significance. Demonstrated on the GHZ state, Fig.~\ref{fig:DS} shows that the learning algorithm scales reasonably with small $d$. The query complexity also scales well with respect to $L$ on average. Such findings give numerical credence to our previous claims about the general efficiency of the learning algorithm, at least for moderately sized systems.

\subsection{\label{sec:results:sub:reg}Regularization Performance}
The heart of the CNet is a classical estimation of $\CL_{\ket{\psi}}(\bth)$. The more rapidly the local landscape of a symmetry submanifold can be fully learned (i.e. estimated to low error) by the CNet, the more efficiently our scheme finds new symmetries. 
Measurement of classical learning efficiency manifests differently in the global and local methods. For each, we define a characteristic \textit{learning time}; for the global scheme, \begin{align}
    \t_{G} & = \frac{1}{N} \sum_{n=1}^{N} \operatorname{argmin}_{t \geq 1} [M(\bth^{(n)}_{0}) \neq M(\bth^{(n)}_{t})]
\end{align}
where $\bth^{(n)}_{t}$ is the symmetry proposed on the $t$th epoch of the $n$th fully-reset search (i.e. the CNet is re-initialized to random weights) starting from a random $\bth$, and $M(\bth)$ is is the symmetry submanifold containing $\bth$. For large $N$, $\t_G$ represents the expected number of epochs of training from Nelder-Mead paths the CNet requires to learn a randomly chosen symmetry manifold. For an ideal regularizer, $\t_G = 1$; for a useless regularizer, $\t_G \to \infty$. 

On the other hand, for the local scheme. As the QNet alternates between optimization and exploration with no random restarts, the key metric of its performance is thus the frequency with which the QNet crosses to a different manifold and back. Hence the learning time may be defined implicitly by \begin{align}
    \t_L(T)^{-1} = \frac{1}{T} \sum_{t=1}^{T} \mathbbm{1}[M(\bth_{t-1}) \neq M(\bth_t)] ,
\end{align}
where $\bth_t$ is the symmetry on the $t$th optimization epoch. For large $T$, $\t_L$ has the same interpretation as $\t_G$.

\begin{figure}[ht!]
    \centering
    \includegraphics[scale=0.5]{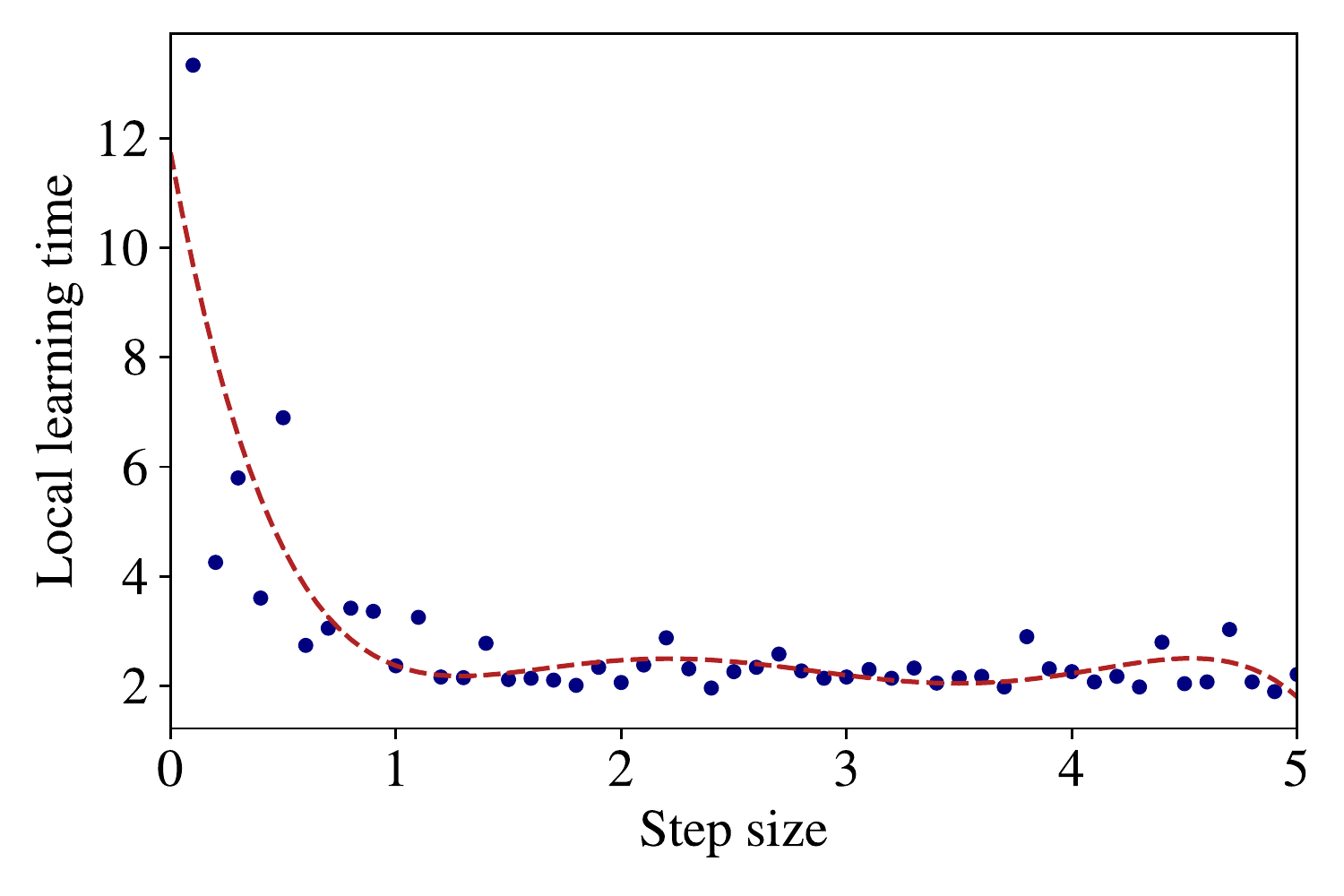}
    \caption{Local learning time $\t_L(400)$ (blue dots) fitted with a trend line (red, dashed) for ease of visualization. $\t_L$ represents the average number of points on a symmetry submanifold needed to be learned before the CNet recognizes the entire submanifold---the smaller $\t_L$ is, the more efficiently the algorithm regularizes. For larger gradient descent step sizes, the learning time saturates.}
    \label{fig:Regularizer}
\end{figure}

Explicit measurement of the learning time requires analytical knowledge of the set of all symmetry submanifolds. Since we have written that set for $S_0[\ket{\mathrm{GHZ}}_L]$, we will measure the learning times for both methods on the $3$-GHZ state. In the global scheme, over 100 runs we found $\t \approx 1.2$ regardless of the steps $N$ per CNet query, in a range of $N = 100$ to $N = 5000$. The learning time of the local method depends on the distance walked in the exploration phase, which can be measured either by the step size in the gradient descent procedure or in the number of iterations; we fix the latter and use the former, and plot the learning time (which can be interpreted as a scoring function since bigger is better) in Fig.~\ref{fig:Regularizer} for $T = 400$, fitted by standard methods in the dashed line to guide the eye. The shift rule also requires a finite difference parameter $h$, but we found that the learning time does not depend significantly on it.
Figure~\ref{fig:Regularizer} illustrates the intuition that, for small step sizes/time spent exploring, the algorithm requires much more time to learn each manifold because it cannot walk sufficiently far away from a symmetry submanifold before the exploratory phase ends. For sufficiently large step size, however, ($\approx 5$ for 3-GHZ), the exploration has taken the QNet sufficiently far away from the known symmetry submanifold that more exploration is unnecessary. Although Fig.~\ref{fig:Regularizer} benchmarks a simple state, the lessons taught by the results hold more generally in that a sufficiently large choice of step size is crucial to build an effective local regularizer.
\subsection{Learning in the Presence of Noise}
A final consideration for implementation in near-term devices is the robustness of the learning algorithm in the presence of noise. For simplicity, we consider a noise model, simulated classically, for which each gate (both in the state preparation and $C_{L, d}$) can incur a bit flip (erroneous $X$ error) or reset (qubit resets to $\ket{0}$), each independently with probability $p_{\text{error}}$. The measurement of each qubit may also be erroneous with the same probability. The presence of such noise may be physically interpreted as thermal fluctuations that jiggle the path traversed by the QNet in $\mathrm{U}$. Because our QNet requires many iterations to propose a symmetry---$10^5$ or more for some systems, as shown in Fig.~\ref{fig:DS}---we expect that such small thermal fluctuations average out in the process, at least up to a certain threshold of $p_{\text{error}}$. 

We examine the loss metric $\CL_{\ket{\psi}}(\bth)$ as a function of $p_{\text{error}}$ on the 3-GHZ state for $d = 1$ to demonstrate this effect numerically.
\begin{figure}[ht!]
    \centering
    \includegraphics[scale=0.5]{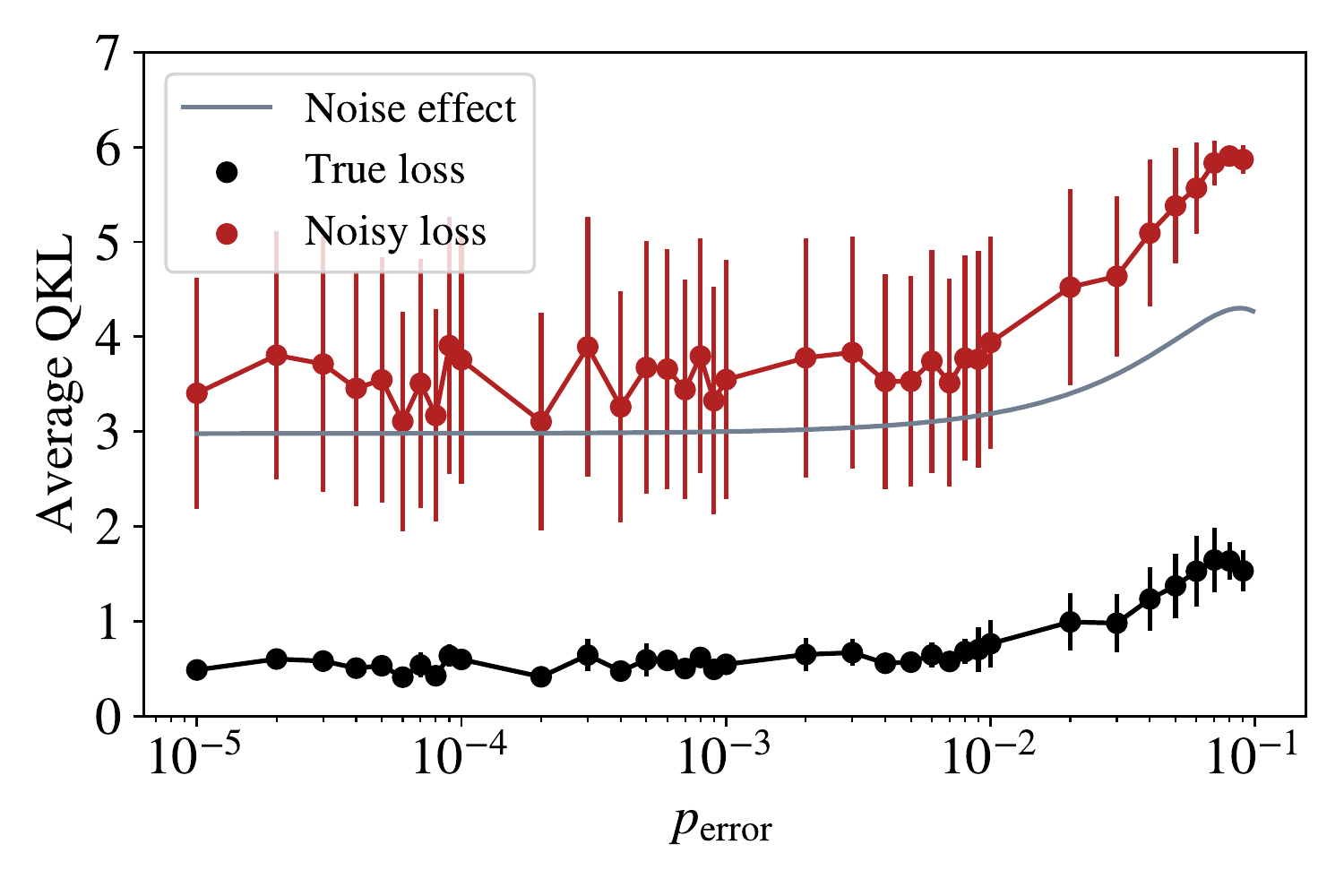}
    \caption{Average loss over $20$ symmetries proposed by a noisy circuit with respect to a noisily prepared GHZ state (red) and a noiseless GHZ state (black), which represents the true loss of the noisy symmetry. Error bars show log-standard deviation. The grey line is a fitted curve of the difference in losses evaluated with respect to a noisy state and a noiseless state.}
    \label{fig:Noise}
\end{figure}
The black curve in Fig.~\ref{fig:Noise} represents the true average loss of the symmetry; that is, the QKL value between the shadows of $U \ket{\mathrm{GHZ}_3}$ and $\ket{\mathrm{GHZ}_3}$, where $U$ is the symmetry proposed by the noisy PQC. Until $p_{\text{error}} \sim 10^{-2}$, the loss remains relatively low and constant. However, in practice we cannot calculate the true loss due to noisiness of state preparation. When $\ket{\mathrm{GHZ}_3}$ is itself is prepared with noise, the resultant average QKL value is given in red. As expected, they are much higher, but the difference between the two, plotted with standard smoothening methods in grey, also remains relatively constant until $p_{\text{error}} \sim 10^{2}$. Thus, when the error is below a reasonably high threshold, the learning algorithm continues to propose true symmetries, and the amount by which the loss increases due to the noisiness of state preparation varies little. By subtracting out this constant gap, we can account for the effect of the noise and thereby continue to learn symmetries in the presence of noise. Above the threshold, the loss increases roughly linearly and slowly, and symmetries may still be classified, albeit slightly noisily.

\section{Conclusion \& Outlook} \label{sec:conclusion}
We developed an interactive hybrid quantum-classical learning algorithm that efficiently solves the symmetry learning problem---to discover and classify every symmetry of an unknown state $\ket{\psi}$ that can be represented by a certain quantum circuit family with a fixed block-depth $d$. We first showed that known symmetries could be verified efficiently via a quantum algorithm, and then devised a method to upgrade the verification procedure into a variational quantum learning algorithm that generated symmetries. To regularize against repetitive propositions of similar symmetries, we introduced an interactive protocol with a classical deep neural network that guided the variational quantum algorithm away from areas of unitary space $\mathrm{U}(2^L)$ explored in previous iterations of learning. Thus, for each state we demonstrated it upon, the algorithm converged. For purposes of near-term implementation, we showed that aspects of the algorithm difficult to implement in current hardware could be replaced by presently realizable methods at the cost of some efficiency. We benchmarked our algorithm in simple cases and showed that it scales well with respect to $d$ and $L$. We further showed that the learning scheme is robust against noise.

Although the parameterized quantum circuit families we discussed are universal or near-universal, a general symmetry may require a very deep circuit in the family to be represented. In practice, under a restriction of depth, we can detect more symmetries by choosing various families of circuits and running the learning algorithm on each. Since some symmetries are more easily represented by certain circuit architectures than others, such an approach maximizes the number of symmetries found empirically on near-term devices.

An important next step in the utilization of our algorithm is more robust benchmarking on various quantum hardware, such as IBM digital quantum devices or Rydberg machines. More analysis on a broader set of states is also of interest. As for regularization, a more complex protocol may allow for even more efficiency. For example, a procedure based on the Metropolis-Hastings algorithm, which stochastically combines the global and local approaches proposed in this paper, is a natural next step of study. Indeed, recent work has shown that application of similar approaches based on Markov chain Monte Carlo methods significantly improve the efficiency of many variational quantum algorithms~\cite{patti2021markov}.

\updated{
Beyond experimental considerations, two major questions remain. The first is physical: while we address the computational problem of symmetry learning, it remains unclear as to how the symmetries are to be physically interpreted. Historically, searching the symmetries of a system generally begins with some intuitive interpretation from which the equations are obtained. If, on the other hand, one has a general matrix or circuit which is known to be a symmetry, it is unclear as to how one would extract the physics from it. The second question is complexity-theoretic: although our work is manifestly computational, it should be possible to derive analytical complexity bounds on interactive quantum-classical search. Numerous investigations of variational search complexities have been conducted in recent years, such as that by \citet{bittel2021training}.
}
In a broader context, symmetry learning can serve as a data-driven guide to learning about the physical phenomena of quantum systems and their relation to symmetries. For example, symmetry breaking is closely related to phase transitions and other emergent properties. Symmetry learning also aligns with the goals of partial tomography in providing a characterization of a state without completely reconstructing it classically. While recent methods such as shadow tomography~\cite{aaronson2019shadow} offer an efficient and convenient way to understand properties of a state from a computational perspective, symmetry learning may be of more relevance from a physics perspective, guiding researchers in quantum machine learning and shedding light on the physics of complex, empirically created quantum systems such as spin liquids~\cite{Semeghini_2021} in addition to boosting other quantum learning algorithms that benefit from knowing symmetries.

\begin{acknowledgments}
We thank Milan Kornjača for helpful discussions on Rydberg atoms. The classical simulations 
in this paper were performed on the FASRC Cannon cluster supported by the FAS
Division of Science Research Computing Group at Harvard University. RAB acknowledges
support from NSF Graduate Research Fellowship under Grant No. DGE1745303, as well as funding from Harvard University's Graduate Prize Fellowship. 
SFY acknowledges funding from NSF and AFOSR.
\end{acknowledgments}

\appendix

\section{\label{app:2bases}QKL Basis Requirement}
The definition of the QKL divergence requires a choice of a number of bases. For every family of states we considered, only 2 constant bases were necessary. We show this by defining a QKL cross validation loss and showing that for 2 constant bases, the cross validation is sufficiently small.

The methodology is as follows. Choose $2$ bases to train on. These may be chosen randomly. However, if one knows that $\ket{\psi}$ has most of its amplitude concentrated in the small proportion of the basis elements in some basis, that basis will be optimal to train on. The second basis may be obtained by perturbing the optimal basis by a small rotation angle. We find that a rotation about an arbitrary axis of $\pi/10$ suffices to capture all of the phase information of the state. 

To verify that 2 bases suffices, choose $n$ random bases after training and evaluate the QKL value over the random set. If it is under the desired error threshold, the two bases suffice. In our case, we found that the QKL over 3 random bases was of order $\sim 10^{-8}$ consistently.

\section{\label{app:hyperparam}Hyperparameters}
We use a error tolerance of $10^{-12}$ for Nelder-Mead search, with a maximum allowed iterations of $10^4$ for $L = 3$ qubits and up to $10^6$ for $L = 15$ qubits. The prediction error threshold for the global regularization scheme is $\d = 10^{-2}$. The CNet uses 100 nodes per layer in the fully connected neural network component. We chose these parameters for their good numerical results and rapid convergence in the systems studied in this paper, but different states or those with sizes larger than the sizes considered in our study may require different choices.

\section{\label{app:cnet}CNet Demonstration}
\begin{figure}[ht!]
    \centering
    \includegraphics[scale=0.5]{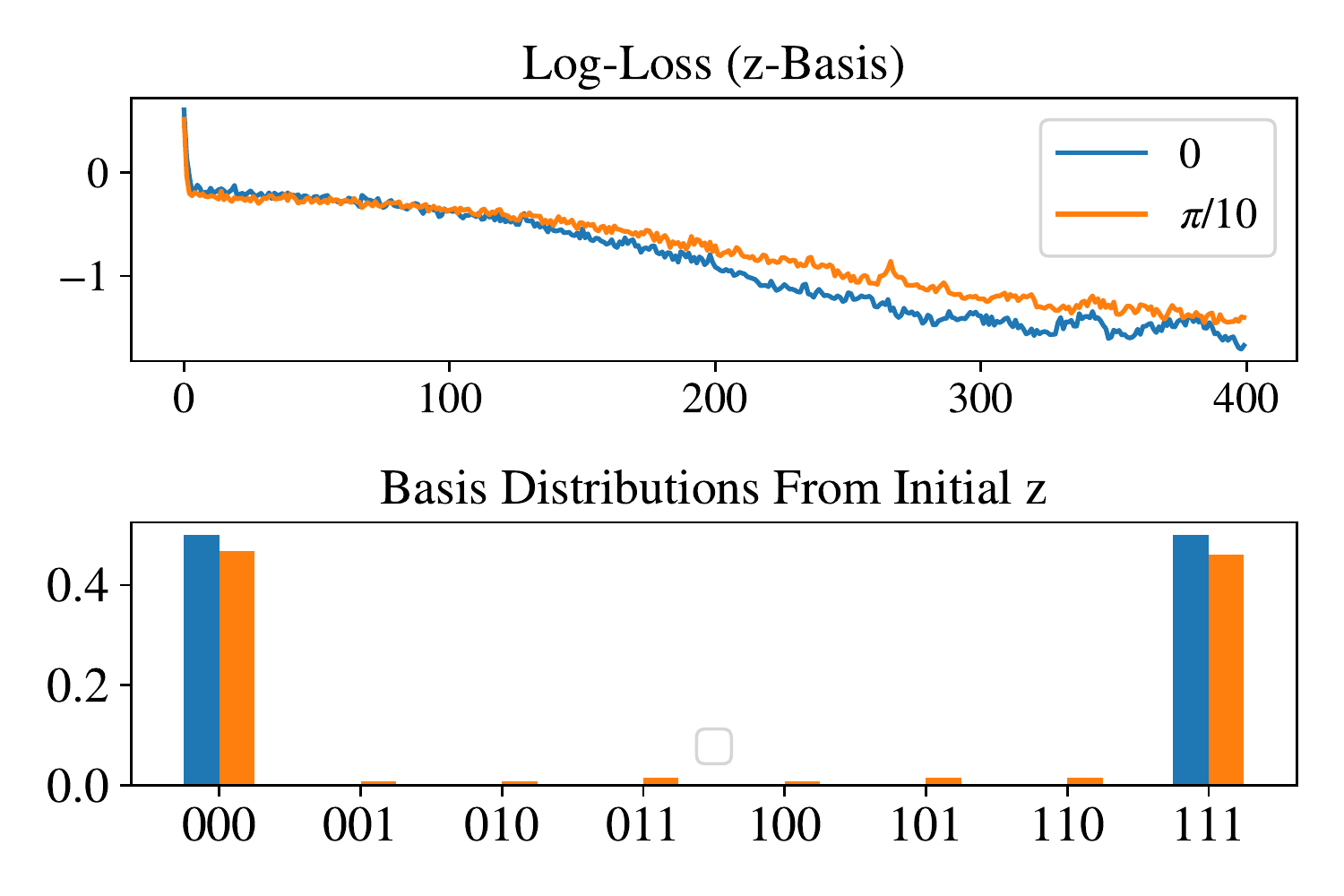}
    \caption{Top: $\log \D_C$ of CNets under each basis ($0$ [blue] corresponds to the $z$-basis). Bottom: visualization of the shadows of the 3-GHZ state.}
    \label{fig:app:CNet_learning}
\end{figure}

It is instructive to examine the classical deep net independently to verify its ability to learn the loss metric landscape. As an example, we consider the CNet on the 3-qubit GHZ state with a block-depth of $d = 2$. We generate a simple dataset of $3000$ random parameters on two bases, one the computational (z) basis and the other a perturbation by a rotation of $\pi/10$ radians about a fixed axis. For each basis, the respective CNet learns the projected classical KL value on the training set, then attempt to estimate it on a testing set of $500$ values. The resulting learning curve is given in Fig.~\ref{fig:app:CNet_learning}.

The computational basis learns to a training error of $0.024$ and incurs a cross-validation (test) error of $1.372$, while the perturbed basis has a training/validation error of $0.038$/$1.224$. The gap between the training and cross-validation error spans two orders of magnitude. Such effects, already visible in this toy example, are dramatically enhanced in practice wherein the training set is a local path rather than a random global sample, for the correlation between points is much higher, leading to a lower training loss and a higher cross-validation error. This is precisely the desired property of the CNet. 

\bibliography{apssamp}

\end{document}